\documentclass[11pt,letterpaper]{article}
\usepackage[centertags]{amsmath}
\usepackage{amssymb}

\usepackage{graphicx}
\usepackage{dcolumn}
\usepackage{bm}
\newcommand{\Z}{{\mathbb Z}}

\newcommand{\drawsquare}[2]{\hbox{%
\rule{#2pt}{#1pt}\hskip-#2pt
\rule{#1pt}{#2pt}\hskip-#1pt
\rule[#1pt]{#1pt}{#2pt}}\rule[#1pt]{#2pt}{#2pt}\hskip-#2pt
\rule{#2pt}{#1pt}}

\newcommand{\fund}{\raisebox{-.5pt}{\drawsquare{6.5}{0.4}}}
\newcommand{\Ysymm}{\raisebox{-.5pt}{\drawsquare{6.5}{0.4}}\hskip-0.4pt%
        \raisebox{-.5pt}{\drawsquare{6.5}{0.4}}}
\newcommand{\Yasymm}{\raisebox{-3.5pt}{\drawsquare{6.5}{0.4}}\hskip-6.9pt%
        \raisebox{3pt}{\drawsquare{6.5}{0.4}}}
\newcommand{\antifund}{\overline{\fund}}

\begin{document}
\pagenumbering{arabic}
\begin{flushright}
\baselineskip=12pt \normalsize
{ACT-08-06}\\
{MIFP-06-30}\\
\smallskip
\end{flushright}

\begin{center}
\Large {\textbf{MSSM via Pati-Salam from Intersecting Branes on $T^6/(\mathbf{\Z_2} \times \mathbf{\Z_2'})$}}        \\[2cm]
\normalsize Ching-Ming Chen$^{\dag\,1}$,  V. E. Mayes$^{\dag\,2}$,
D. V. Nanopoulos$^{\dag\, \ddag\, \S\,3}$
\\[.25in]
\textit{$^\dag$George P. and Cynthia W. Mitchell Institute for
Fundamental Physics, \\Texas A$\&$M University, College Station,
TX 77843, USA
\\
$^\ddag$Astroparticle Physics Group, Houston Advanced Research
Center (HARC), \\Mitchell Campus, Woodlands, TX 77381, USA
\\
$^\S$Academy of Athens, Division of Natural Sciences, \\28
Panepistimiou Avenue, Athens 10679, Greece} \\[.5cm]
\tt \footnotesize $^1$cchen@physics.tamu.edu,
$^2$eric@physics.tamu.edu,\\
$^3$dimitri@physics.tamu.edu\\[2cm]
\end{center}
\begin{abstract} 
We construct an MSSM-like model via Pati-Salam from intersecting D-branes in Type IIA theory
on the $\Z_2 \times \Z_2'$ orientifold where the D-branes wrap rigid 3-cycles. Because the 3-cycles are rigid, there
are no extra massless fields in the adjoint representation, arising as open-string moduli. The presence of these unobserved fields would create difficulties with asymptotic freedom as well as the prediction of gauge unification. The model constructed has four generations of MSSM matter plus right-handed neutrinos, as well as additional vector-like representations. In addition, we find that all of the required Yukawa couplings are allowed by global symmetries which arise from $U(1)$'s which become massive via a generalized Green-Schwarz mechanism.  Furthermore, we find that the tree-level gauge couplings are unified at
the string scale. 
\end{abstract}

\section{Introduction}
At the present, string theory is the only framework for realization of a unification of gravitation with gauge theory and quantum mechanics.  In principle, it should be possible to derive all known physics from the string, as well as potentially provide something new and unexpected.  This is the goal of string phenomenology.  However, in spite of this there exist many solutions that may be derived from string, all of which are consistent vacua. One of these vacua should correspond to our universe, but then the question becomes why this
particular vacuum is selected.  One possible approach to this state of affairs is to statistically classify the possible vacua, in essence making a topographical map of the \lq landscape\rq.  One then attempts to assess the liklihood that vacua with properties similar to ours will arise.\footnote{For example, see~\cite{Blumenhagen:2004xx, Gmeiner:2005vz, Douglas:2006xy, Gmeiner:2006qw, Gmeiner:2006vb,Dienes:2006ut}.}  Another approach is to take the point of view that there are unknown dynamics, perhaps involving a departure from criticality, which determine the vacuum that corresponds to our universe.  Regardless of the question of uniqueness, if string theory is correct then it should be possible to find a solution
which corresponds \textit{exactly} to our universe, at least in it's low energy limit.  Although there has been a great deal of progress in constructing
semi-realistic models, this has not yet been achieved. 

An elegant approach to model construction involving
Type I orientifold (Type II) compactifications is 
where chiral fermions  arise from strings
stretching between D-branes intersecting at angles (Type IIA
picture) \cite{Berkooz:1996km} or in its T-dual (Type IIB)
picture with magnetized D-branes \cite{Bachas:1995ik}.
Many consistent standard-like and grand unified theory (GUT)
models have been constructed~\cite{Blumenhagen:2000wh,
Aldazabal:2000dg, Angelantonj:2000hi, Ellis:2002ci} using D-brane
constructions.  The first quasi-realistic supersymmetric models
were constructed in Type IIA theory on a $\mathbf{T^6} /(\Z_2
\times \Z_2)$ orientifold~\cite{CveticShiuUranga, Cvetic:2001tj}.
Following this, models with standard-like, left-right symmetric
(Pati-Salam), unflipped $SU(5)$ gauge
groups were constructed based upon the same framework and
systematically studied \cite{Cvetic:P:S:L,Cvetic:2002pj, Cvetic:2004nk,Chen:2006gd}. In addition, several different flipped $SU(5)$~\cite{Barr:1981qv,FSU(5)N,AEHN} models have also been built using intersecting D-brane constructions~
\cite{Ellis:2002ci, Chen:2005ab, Axenides:2003hs,Chen:2005cf,Chen:2005mm,Chen:2006ip}.\footnote{For excellent reviews, see~\cite{Blumenhagen:2005mu} and~\cite{Blumenhagen:2006ci}.}

Although much progress has been made, none of these models have been completely satisfactory.  Problems include extra chiral and non-chiral matter, and the lack
of a complete set of Yukawa couplings, which are typically forbidden by global symmetries. In addition to the chiral matter which arises at brane intersections, D-brane constructions typically will have non-chiral open string states present in the low-energy spectrum associated with the D-brane position in the internal space and Wilson lines.  This results in adjoint or additional matter in the
symmetric and antisymmetric representations unless the open string moduli are completely frozen.  These light scalars are not observed and are not present in the MSSM.  While it is possible that these moduli will obtain  mass after supersymmetry is broken, it would typically be of the TeV scale.  While this would make them unobservable in present experiments, the succesful gauge unification in the MSSM would be spoiled by their presence.  While it may be possible to find some scenarios where the problems created by these fields are ameliorated, it is much simpler to eliminate these fields altogether.  One way to do this is to this is to construct intersecting D-brane models where the D-branes wrap rigid cycles.\footnote{This possibility was first explored in~\cite{Dudas:2005jx} and~\cite{Blumenhagen:2005tn}.} Another motiviation for the absence of
these adjoint states is that this is consistent with a $k=1$ Kac-Moody algebra in models constructed from heterotic string, some
of which may be dual.

In this letter, we construct an intersecting D-brane model on the $\Z_2 \times \Z_2'$ orientifold background, also known as the $\Z_2 \times \Z_2$ orientifold with discrete torsion, where the D-branes wrap rigid cycles thus eliminating the extra adjoint matter.  This letter is organized as follows:  First, we briefly review intersecting D-brane constructions on the $\Z_2 \times \Z_2'$ orientifold. We then proceed to construct a supersymmetric four-generation MSSM-like model obtained from a Pati-Salam model via spontaneous gauge symmetry breaking. All of the required Yukawa couplings are allowed by global symmetries present in the movel.  We find that the tree-level gauge couplings are unified at
the string scale. 

\section{Intersecting Branes on the $\Z_2 \times \Z_2$ Orientifold with and without Discrete Torsion}
The $\Z_2 \times \Z_2$ orientifold has been the subject of extensive research, primarily because it is the simplest background space which allows supersymmetric vacua.  We will essentially follow along with the development given in~\cite{Blumenhagen:2005tn}.  
The first supersymmetric models based upon the $\Z_2 \times \Z_2$ orientifold were explored in ~\cite{CveticShiuUranga,Cvetic:2001tj,Cvetic:P:S:L,Cvetic:2002pj}. In Type IIA theory on the $\mathbf{T^6} /(\Z_2 \times \Z_2)$ orientifold, the $\mathbf{T^6}$ is product of three
two-tori and the two orbifold group generators $\theta$, $\omega$
act on the complex coordinates $(z_1,z_2,z_3)$ as
\begin{eqnarray}
\theta:(z_1,z_2,z_3)\rightarrow(-z_1,-z_2,z_3) \nonumber \\
\omega:(z_1,z_2,z_3)\rightarrow(z_1,-z_2,-z_3)
\end{eqnarray}
while the antiholomorphic involution $R$ acts as 
\begin{equation}
R(z_1, z_2, z_3)\rightarrow(\bar{z}_1,\bar{z}_2,\bar{z}_3).
\end{equation}
As it stands, the signs of the $\theta$ action in the $\omega$ sector and vice versa have not been specified, and the freedom to do so is referred to as the choice of discrete torsion.  One choice of discrete torsion corresponds to the Hodge numbers $(h_{11},h_{21}) = (3,51)$ and the corresponding to $(h_{11},h_{21}) = (51,3)$.  These two different choices are referred to as with discrete torsion $(\Z_2 \times \Z_2')$ and without discrete torsion $(\Z_2 \times \Z_2)$ respectively.  To date, most phenomenological models that have been constructed have been without discrete torsion.  Consequently, all of these models have massless adjoint matter present since the D-branes do not wrap rigid 3-cycles.  However, in the case of $\Z_2 \times \Z_2'$, the twisted homology contains collapsed 3-cycles, which allows for the construction of rigid 3-cycles. 

D6-branes wrapping cycles are specified by their wrapping numbers $(n^i, m^i)$
along the fundamental cycles $[a^i]$ and $[b^i]$ on each torus.  However, cycles on the torus are, in general, different from the cycles defined on the orbifold space.  In the case of the $\Z_2 \times \Z_2$ orientifold, all of the 3-cycles on the orbifold are inherited from the torus, which makes it particulary easy to work with.  
The $\Z_2 \times \Z_2'$ orientifold contains 16 fixed points, from which arise 16 additional 2-cycles with the topology of $\mathbf{P}^1 \cong S^2$.  As a result, there are 32 collapsed 3-cycles for each twisted sector.  A $D6$-brane wrapping collapsed 3-cycles in each of the three twisted sectors will be unable to move away from a particular position on the covering space $\mathbf{T^6}$, which means that the 3-cycle will be rigid.    

A basis of twisted 3-cycles may be denoted as
\begin{eqnarray}
[\alpha^{\theta}_{ij,n}] &=& 2[\epsilon^{\theta}_{ij}]\otimes [a^3] \ \ \ \ \ \ \ \ \  [\alpha^{\theta}_{ij,m}] = 2[\epsilon^{\theta}_{ij}]\otimes [b^3],
\end{eqnarray}
\begin{eqnarray}
[\alpha^{\omega}_{ij,n}] &=& 2[\epsilon^{\omega}_{ij}]\otimes [a^1] \ \ \ \ \ \ \ \ \  [\alpha^{\omega}_{ij,m}] = 2[\epsilon^{\omega}_{ij}]\otimes [b^1],
\end{eqnarray}
\begin{eqnarray}
[\alpha^{\theta\omega}_{ij,n}] &=& 2[\epsilon^{\theta\omega}_{ij}]\otimes [a^2] \ \ \ \ \ \ \ \ \  [\alpha^{\theta\omega}_{ij,m}] = 2[\epsilon^{\theta\omega}_{ij}]\otimes [b^2].
\end{eqnarray}
where $[\epsilon^{\theta}_{ij}]$, $[\epsilon^{\omega}_{ij}]$, and $[\epsilon^{\theta\omega}_{ij}]$ denote the 16 fixed points on $\mathbf{T}^2 \times  \mathbf{T}^2$, where $i,j \in {1,2,3,4}$. 

A fractional D-brane wrapping both a bulk cycle as well as the collapsed cycles may be written in the form
\begin{eqnarray}
\Pi^F_a &=& \frac{1}{4}\Pi^B + \frac{1}{4}\left(\sum_{i,j\in S^a_{\theta}} \epsilon^{\theta}_{a,ij}\Pi^{\theta}_{ij,a}\right)+ \frac{1}{4}\left(\sum_{j,k\in S^a_{\omega}} \epsilon^{\omega}_{a,jk}\Pi^{\omega}_{jk,a}\right)
+ \frac{1}{4}\left(\sum_{i,k\in S^a_{\theta\omega}} \epsilon^{\theta\omega}_{a,ik}\Pi^{\theta\omega}_{ik,a}\right).
\label{fraccycle}
\end{eqnarray}
where the $D6$-brane is required to run through the four fixed points for each of the twisted sectors.  The set of four fixed points may be denoted as $S^g$ for the twisted sector $g$. The constants $\epsilon^{\theta}_{a,ij}$, $\epsilon^{\omega}_{a,jk}$ and $\epsilon^{\theta\omega}_{a,ki}$ denote the sign of the charge of the fractional brane with respect to the fields which are present at the orbifold fixed points.  These signs, as well as the set of fixed points, must satisfy consistency conditions.  However, they may be chosen differently for each stack.  

A bulk cycle on the $\Z_2 \times \Z_2$ orbifold space consist of the toroidal cycle wrapped by the brane $D_a$ and it's three orbifold images:
\begin{eqnarray}
\left[\Pi^B_a \right] &=& \left(1 + \theta + \omega + \theta\omega \right)\Pi^{T^6}_a. 
\end{eqnarray}
Each of these orbifold images in homologically identical to the original cycle, thus
\begin{eqnarray}
\left[\Pi^B_a \right] &=& 4\left[\Pi^{T^6}_a \right].
\end{eqnarray}
If we calculate the intersection number between two branes, we will find
\begin{eqnarray}
\left[\Pi^B_a\right] \circ \left[\Pi^B_b\right] &=& 4~\left[\Pi^{T^6}_a \right]\circ \left[\Pi^{T^6}_b\right]
\end{eqnarray}
which indicates that the bulk cycles $\left[\Pi^B_a\right]$ do not expand a unimodular basis for the homology lattice $H_3(M,Z)$.  Thus, we must normalize these purely bulk cycles as
$\left[\Pi^o_a\right] = \frac{1}{2}\left[\Pi^B_a\right]$~\cite{Blumenhagen:2005mu, Blumenhagen:2005tn}. So, in terms of the cycles defined on the torus, the normalized purely bulk cycles of the orbifold are given by
\begin{eqnarray}
\left[\Pi^o_a\right] &=& \frac{1}{2}\left(1 + \theta + \omega + \theta\omega \right)\left[\Pi^{T^6}_a\right] = 2\left[\Pi^{T^6}_a\right].
\label{eqn:orbthreecycle}
\end{eqnarray}
Due to this normalization, a stack of $N$ $D6$-branes wrapping a purely bulk cycle will have a $U(N/2)$ gauge group
in its world-volume.  However, this does not apply to a brane wrapping collapsed cycles, so that a stack of $N$ branes wrapping fractional cycles as in eq.~\ref{fraccycle} will have in its world-volume a gauge group $U(N)$.  

Since we will have $D6$-branes which are wrapping fractional cycles with a bulk component as well as twisted cycles,
we will need to be able to calculate the intersection numbers between pairs of twisted 3-cycles.  
For the intersection number between two twisted 3 cycles of the form $[\Pi^g_{ij,a}] = n^{I_g}_a[\alpha_{ij,n}]+m^{I_g}_a[\alpha_{ij,m}]$ and $[\Pi^h_{kl,b}] = n^{I_h}_b[\alpha_{kl,n}]+m^{I_h}_b[\alpha_{kl,m}]$ we have
\begin{eqnarray}
[\Pi^g_{ij,a}] \circ [\Pi^h_{kl,b}] &=& 4\delta_{ik}\delta_{jl}\delta^{gh}(n^{I_g}_am^{I_g}_b - m^{I_g}_a n^{I_g}_b)
\end{eqnarray}
where $I_g$ corresponds to the torus left invariant by the action of the orbifold generator $g$; specifically $I_{\theta} = 3$, $I_{\omega} = 1$, and $I_{\theta\omega} = 2$.  

Putting everything together, we will find for the intersection number between a brane $a$ and brane $b$ wrapping fractional cycles we will have
\begin{eqnarray}
\Pi^F_a \circ \Pi^F_b = \frac{1}{16}[\Pi^B_a \circ \Pi^B_b + 4(n_a^3m_b^3-m_a^3n_b^3)\sum_{i_aj_a\in S^a_{\theta}}\sum_{i_bj_b\in S^b_{\theta}}\epsilon^{\theta}_{a,i_aj_a}\epsilon^{\theta}_{b,i_bj_b}\delta_{i_ai_b}\delta_{j_aj_b} 
+ \\ \nonumber 4(n_a^1m_b^1-m_a^1n_b^1)\sum_{j_ak_a\in S^a_{\omega}}\sum_{j_bk_b\in S^b_{\omega}}\epsilon^{\omega}_{a,j_ak_a}\epsilon^{\omega}_{b,j_bk_b}\delta_{j_aj_b}\delta_{k_ak_b}
+ \\ \nonumber
4(n_a^2m_b^2-m_a^2n_b^2)\sum_{i_ak_a\in S^a_{\theta\omega}}\sum_{i_bk_b \in S^b_{\theta\omega}}\epsilon^{\theta\omega}_{a,i_ak_a}\epsilon^{\theta\omega}_{b,i_bk_b}\delta_{i_ai_b}\delta_{k_ak_b}].
\end{eqnarray}

The 3-cycle wrapped by the $O6$-planes is given by 
\begin{equation}
2q_{\Omega R}[a^1][a^2][a^3]-2q_{\Omega R\theta}[b^1][b^2][a^3]-2q_{\Omega R\omega}[a^1][b^2][b^3]-2q_{\Omega R\theta\omega}[b^1][a^2][b^3].  
\end{equation}
where the cross-cap charges $q_{\Omega R g}$ give the RR charge and tension of a given orientifold plane $g$, of which there are two
types, $O6^{(-,-)}$ and $O6^{(+,+)}$.  In this case, $q_{\Omega R g} = +1$ indicates an $O6^{(-,-)}$ plane, while 
$q_{\Omega R g} = -1$ indicates an $O6^{(+,+)}$ while the choice of discrete torsion is indicated by the product
\begin{equation}
q = \prod_g q_{\Omega R g}.
\end{equation}
The choice of no discrete torsion is given by $q = 1$, while for $q = -1$ is the case of discrete torsion, for which
an odd number of $O^{(+,+)}$ must be present.  

The action of $\Omega R$ on the bulk cycles is the same in either case, and is essentially just changes the signs of the wrapping numbers as $n^i_a \rightarrow n^i_a$ and $m^i_a \rightarrow -m^i_a$.  However, in addition, there is an action
on the twisted 3 cycle as 
\begin{eqnarray}
\alpha^g_{ij,n} \rightarrow -q_{\Omega R}q_{\Omega Rg}\alpha^g_{ij,n}, & \alpha^g_{ij,m} \rightarrow q_{\Omega R}q_{\Omega Rg}\alpha^g_{ij,m}.
\end{eqnarray}
Using these relations, one can work out the intersection number of a fractional cycle with it's $\Omega R$ image, we have
\begin{eqnarray}
\Pi'^F_a \circ \Pi^F_a = q_{\Omega R}\left(2q_{\Omega R}\prod_In^I_am^I_a - 2q_{\Omega R\theta}n^3_am^3_a 
-2q_{\Omega R\omega}n^1_am^1_a - 2q_{\Omega R\theta\omega}n^2_am^2_a\right)
\end{eqnarray}
while the intersection number with the orientifold planes is given by
\begin{eqnarray}
\Pi_{O6} \circ \Pi^F_a = 2q_{\Omega R}\prod_I m^I_a - 2q_{\Omega R\theta}n^1_an^2_am^3_a - 2q_{\Omega R\omega}m^1_an^2_an^3_a - 2q_{\Omega R\theta\omega}n^1_am^2_an^3_a.
\end{eqnarray}

\begin{table}[f]
\begin{center}
\begin{tabular}{|c|c|c|c|c|c|c|} \hline
         &                       \\
Representation & Multiplicity \\ \hline \hline
         &                       \\
$\mathbf{\Yasymm}$  & $\frac{1}{2}(\left[\Pi^o_{a'} \right]\circ \left[\Pi^o_a\right] + \left[\Pi_{O6}\right] \circ \left[\Pi^o_a\right])$\\  
         &                       \\
$\mathbf{\Ysymm}$ & $\frac{1}{2}(\left[\Pi^o_{a'}\right] \circ \left[\Pi^o_a\right] - \left[\Pi_{O6}\right] \circ \left[\Pi^o_a)\right]$\\ 
         &                       \\
$(\mathbf{\antifund_a}, \mathbf{\fund_b})$ & $\left[\Pi^o_a\right] \circ \left[\Pi^o_b\right]$\\ 
         &                       \\
$(\mathbf{\fund_a}, \mathbf{\fund_b})$ & $\left[\Pi^o_{a'}\right]\circ \left[\Pi^o_{b}\right]$\\ \hline
\end{tabular}
\end{center}
\caption{Net chiral matter spectrum in terms of three-cycles.}
\label{chiralmatter}
\end{table}

A generic expression for the \textit{net} number of chiral fermions in bifundamental, symmetric, and antisymmetric representations consistent with the vanishing of RR tadpoles can be given in terms of the three-cycles cycles~\cite{Blumenhagen:2002wn} which is shown in Table \ref{chiralmatter}.  
\section{Consistency and SUSY conditions}
Certain conditions must be applied to construct consistent, supersymmetric vacua which are free of anomalies, which
we discuss in the following sections. 
\subsection{RR and Torsion Charge Cancellation}
With the choice of discrete torsion $q_{\Omega R} = -1$, $q_{\Omega R\theta} = q_{\Omega R\omega} = q_{\Omega R\theta\omega} = 1$, the condition for the cancellation of RR tadpoles becomes
\begin{eqnarray}
\sum N_a n_a^1 n_a^2 n_a^3 = -16, & 
\sum N_a m_a^1 m_a^2 n_a^3 = -16, \\ \nonumber
\sum N_a m_a^1 n_a^2 m_a^3 = -16, &
\sum N_a n_a^1 m_a^2 m_a^3 = -16.
\end{eqnarray}
whilst for the twisted charges to cancel, we require 
\begin{eqnarray}
\sum_{a, ij \in S^{\omega}} N_a n_a^1 \epsilon^{\omega}_{ij,a} = 0, \ \ \ \ 
\sum_{a, jk \in S^{\theta\omega}} N_a n_a^2 \epsilon^{\theta\omega}_{jk,a} = 0, \ \ \ \ 
\sum_{a, ki \in S^{\theta}} N_a n_a^3 \epsilon^{\theta}_{ki,a} = 0.
\end{eqnarray}
where the sum is over \textit{each} each fixed point $[e^g_{ij}]$.  
As stated in Section 2, the signs $\epsilon^{\theta}_{ij,a}$, $\epsilon^{\omega}_{jk,a}$, and $\epsilon^{\theta\omega}_{ki,a}$ are not arbitrary as they must satisfy certain consisitency conditions.  In particular, they must satisfy the condition
\begin{eqnarray}
\sum_{ij \in S^g}\epsilon^g_{a,ij} = 0 \ \ \ \mbox{mod} \ \ \ 4
\end{eqnarray}
for each twisted sector.  Additionally, the signs for different twisted sectors must satisfy
\begin{eqnarray}
\epsilon^{\theta}_{a,ij}\epsilon^{\omega}_{a,jk}\epsilon^{\theta\omega}_{a,ik} &=& 1, \\
\nonumber
\epsilon^{\theta}_{a,ij}\epsilon^{\omega}_{a,jk} &=& \ \mbox{constant} \  \forall \ j.
\end{eqnarray}
Note that we may choose the set of signs differently for each stack provided that they satisfy the consistency
conditions.  A trivial choice of signs which satisfies the constraints placed on them is just to have them all set to $+1$,  
\begin{equation}
\epsilon^{\theta}_{a,ij} = 1 \ \forall \ ij, \ \ \ \ \ \epsilon^{\omega}_{a,jk} = 1 \ \forall \ jk, \ \ \ \ \ \epsilon^{\theta\omega}_{a,ki} = 1\ \forall \ ki. 
\end{equation}
Another possible non-trivial choice of signs consistent with the constraints is given by
\begin{equation}
\epsilon^{\theta}_{a,ij} = -1 \ \forall \ ij, \ \ \ \ \ \epsilon^{\omega}_{a,jk} = -1 \ \forall \ jk, \ \ \ \ \ \epsilon^{\theta\omega}_{a,ki} = 1\ \forall \ ki.
\end{equation}
More general sets of these signs may be found in~\cite{Blumenhagen:2005tn}.  

\subsection{Conditions for Preserving $N=1$ Supersymmetry}

The condition to preserve $\emph{N}=1$ supersymmetry in four
dimensions is that the rotation angle of any D-brane with respect 
to the orientifold plane is an element of $SU(3)$ \cite{Berkooz:1996km,CveticShiuUranga}. Essentially, this becomes a constraint on the 
angles made by each stack of branes with respect to the
orientifold planes, \textit{viz}
$\theta^1_a + \theta^2_a + \theta^3_a = 0$ mod $2\pi$, or equivalently  
$\sin(\theta^1_a
+
  \theta^2_a + \theta^3_a)= 0$ and $\cos(\theta^1_a +
  \theta^2_a + \theta^3_a)= 1$.
Applying simple trigonometry, these angles may be expressed in terms of
the wrapping numbers as
\begin{eqnarray}
\tan \theta^i_a=\frac{m^i_a R^i_2}{n^i_a R^i_1} 
\end{eqnarray}
where $R^i_2$ and $R^i_1$ are the radii of the $i^{\mathrm{th}}$
torus.  We may translate these conditions into restrictions on 
the wrapping numbers as
\begin{eqnarray}
x_A\tilde{A_a}+x_B\tilde{B_a}+x_C\tilde{C_a}+x_D\tilde{D_a}= 0
\nonumber \\
A_a/x_A + B_a/x_B + C_a/x_C + D_a/x_D < 0
\label{susycond}
\end{eqnarray}
where we have made the definitions
\begin{eqnarray}
\tilde{A_a} &=& - m^1_am^2_am^3_a, \ \ \ \tilde{B}_a = n^1_an^2_am^3_a, \ \ \ \tilde{C}_a = m^1_an^2_an^3_a, \ \ \ \tilde{D}_a = n^1_am^2_an^3_a, \\
A_a &=& -n^1_an^2_an^3_a, \ \ \ B_a = m^1_am^1_an^3_a, \ \ \ C_a = n^1_am^1_am^3_a, \ \ \  D_a = m^1_an^1_am^3_a.
\end{eqnarray}
and the structure parameters related to the complex structure moduli are
\begin{eqnarray}
x_a = \lambda,  \ \ \  x_b = \frac{\lambda}{\chi_2\cdot\chi_3}, \ \ \ x_c = \frac{\lambda}{\chi_1\cdot\chi_3}, \ \ \ \frac{\lambda}{\chi_1\cdot\chi_2}.  
\end{eqnarray}
where $\lambda$ is a positive constant.  One may invert the above expressions to find values for the complex structure
moduli as
\begin{eqnarray}
\chi_1 = \lambda, \ \ \ \chi_2 = \frac{x_c}{x_b}\cdot\chi_1, \ \ \ \chi_3 = \frac{x_d}{x_b}\cdot\chi_1.
\end{eqnarray}

\subsection{The Green-Schwarz Mechanism}
Although the total non-Abelian anomaly cancels automatically when
the RR-tadpole conditions are satisfied, additional mixed
anomalies like the mixed gravitational anomalies which generate
massive fields are not trivially zero \cite{CveticShiuUranga}. 
 These anomalies are cancelled by a generalized
Green-Schwarz (G-S) mechanism which involves untwisted
Ramond-Ramond forms. Integrating the G-S couplings of the
untwisted RR forms to the $U(1)$ field strength $F_a$ over the
untwisted cycles of $\mathbf{T^6/(\Z_2\times \Z'_2)}$ orientifold, we find
\begin{eqnarray}
\int_{D6^{untw}_a} C_5 \wedge \textrm{tr}F_a \sim N_a \sum_i
r_{ai}\int_{M_4} B^i_2 \wedge \textrm{tr}F_a,
\end{eqnarray}
where
\begin{equation}
B^i_2 = \int_{[\Sigma_i]} C_5,\;\; [\Pi_a]=\sum^{b_3}_{i=1}
r_{ai}[\Sigma_i],
\end{equation}
and
${[\Sigma_i]}$ is the basis of homology 3-cycles, $b_3=8$. Under orientifold action 
only half survive.  In other words,
$\{r_{ai}\}=\{\tilde{B}_a, \tilde{C}_a, \tilde{D}_a,
\tilde{A}_a\}$ in this definition. Thus the couplings of the four
untwisted RR forms $B^i_2$ to the $U(1)$ field strength $F_a$ are
\cite{Aldazabal:2000dg}
\begin{eqnarray}
  N_a \tilde{B}_a \int_{M_4}B^1_2\wedge \textrm{tr}F_a,&&  \;
  N_a \tilde{C}_a \int_{M_4}B^2_2\wedge \textrm{tr}F_a,
   \nonumber \\
  N_a \tilde{D}_a \int_{M_4}B^3_2\wedge \textrm{tr}F_a,&&  \;
  N_a \tilde{A}_a \int_{M_4}B^4_2\wedge \textrm{tr}F_a.
\end{eqnarray}

Besides the contribution to G-S mechanism from untwisted 3-cycles,
the contribution from the twisted cycles should be taken into
account. As in the untwisted case we integrate the Chern-Simons coupling over the
exceptional 3-cycles from the twisted sector.  We choose the sizes
of the 2-cycles on the topology of $S^2$ on the orbifold
singularities to make the integrals on equal foot to those from
the untwisted sector. Consider the twisted sector $\theta$ as
an example,
\begin{eqnarray}
\int_{D6^{tw,\theta}_a}C_5\wedge {\rm tr}F_a \sim   N_a
\sum_{i,j\in S^a_{\theta}} \epsilon^{\theta}_{a,ij} m^3_a
\int_{M_4} B^{\theta ij}_2 \wedge {\rm tr}F_a.
\end{eqnarray}
where $B^{\theta ij}_2=\int_{[\alpha^{\theta}_{ij,m}]}C_5$, with
orientifold action taken again. Although $i,j$ can run through 
each run through $\left\{1-4\right\}$ for each
of the four fixed points in each sector, these are constrained
by the wrapping numbers from the untwisted sector so that only four
possibilities remain.  A similar argument may be applied for $\omega$ and 
$\theta\omega$ twisted sectors:
\begin{eqnarray}
\int_{D6^{tw,\omega}_a}C_5\wedge {\rm tr}F_a \sim   N_a
\sum_{j,k\in S^a_{\omega}} \epsilon^{\omega}_{a,jk} m^1_a
\int_{M_4} B^{\omega jk}_2 \wedge {\rm tr}F_a.
\end{eqnarray}
\begin{eqnarray}
\int_{D6^{tw,\theta\omega}_a}C_5\wedge {\rm tr}F_a \sim   N_a
\sum_{i,j\in S^a_{\theta\omega}} \epsilon^{\theta\omega}_{a,ik}
m^2_a \int_{M_4} B^{\theta\omega ik}_2 \wedge {\rm tr}F_a.
\end{eqnarray}

In summary, there are twelve additional couplings of the
Ramond-Ramond 2-forms $B^i_2$ to the $U(1)$ field strength $F_a$
from the twisted cycles, giving rise to massive $U(1)$'s.  However
from the consistency condition of the $\epsilon$'s (see section
3.1) related  to the
discrete Wilson lines they may be dependent or degenerate.  So
even including the couplings from the untwisted sector we still
have an opportunity to find a linear combination for a massless
$U(1)$ group.  Let us write down these couplings of the twisted
sector explcitly:
\begin{eqnarray}
N_a  \epsilon^{\theta}_{a,ij} m^3_a
\int_{M_4} B^{\theta ij}_2 \wedge {\rm tr}F_a, \ \ \ 
N_a  \epsilon^{\omega}_{a,jk} m^1_a
\int_{M_4} B^{\omega jk}_2 \wedge {\rm tr}F_a, \nonumber \\
N_a  \epsilon^{\theta\omega}_{a,ik} m^2_a \int_{M_4}
B^{\theta\omega ik}_2 \wedge {\rm tr}F_a.
\end{eqnarray}

Checking the mixed cubic anomaly by introducing the dual field of
$B^i_2$ in the diagram, we can find the contribution from both
untwisted and twisted sectors having a intersection number form
and which is cancelled by the RR-tadpole conditions mentioned.
These couplings determine the linear combinations of $U(1)$ gauge
bosons that acquire string scale masses via the G-S mechanism.  Thus, in constructing
MSSM-like models, we
must ensure that the gauge boson of the hypercharge $U(1)_Y$ group
does not receive such a mass. In general, the hypercharge is a linear combination
of the various $U(1)$s generated from each stack :
\begin{equation}
U(1)_Y=\sum_a c_a U(1)_a
\end{equation}
The corresponding field strength must be orthogonal to those that
acquire G-S mass.  Thus we demand 
\begin{eqnarray}
\sum_a c_a N_a 
\epsilon^{\omega}_{a,jk} m^1_a&=& 0, \ \ \ \ 
\sum_a c_a N_a
\epsilon^{\theta\omega}_{a,ki} m^2_a = 0, \ \ \ \ 
\sum_a c_a N_a
\epsilon^{\theta}_{a,ij} m^3_a = 0, 
\end{eqnarray}
for the twisted couplings as well as 
\begin{eqnarray}
 \sum_a c_a N_a \tilde{A_a} &=& 0, \ \ \ \ 
 \sum_a c_a N_a \tilde{B_a}  = 0, \ \ \ \ 
 \sum_a c_a N_a \tilde{C_a}  = 0, \ \ \ \ 
 \sum_a c_a N_a \tilde{D_a}  = 0, 
\label{GSeq}
\end{eqnarray}
for the untwisted.  

\subsection{K-Theory Constraints}
RR charges are not fully classified by homological data, but rather by K-theory.  Thus, to cancel all charges including those visible by K-theory alone, we require the wrapping numbers to satisfy certain constraints.  We will not state these constraints here, but we will refer the reader to~\cite{Blumenhagen:2005tn} where they are
given explicitely.  

\section{MSSM via Pati-Salam}
We begin with the seven-stack configuration of D-branes with the bulk
wrapping numbers shown in Table~\ref{stacksPS}, which produce the intersection numbers shown in Tables 3-4. We make the choice of cross-cap charges $q_{\Omega R} = -1$, $q_{\Omega R\theta} = q_{\Omega R\omega} = q_{\Omega R\theta\omega} = 1$, and assume for simplicity that each stack passes throught the same set of fixed points.  The resulting gauge group is that of a four generation Pati-Salam model. The \lq observable\rq \ matter spectrum is presented in Table 5.

\begin{table}[f]
\begin{center}
\begin{tabular}{|c|c|c|c|c|c|c|c|c|} \hline
                 Stack & N & $(n_1,m_1)$&($n_2,m_2)$&$(n_3,m_3)$ & $\epsilon^{\theta}_{ij}~\forall~ij$ &  $\epsilon^{\omega}_{jk}~\forall~jk$ & $\epsilon^{\theta\omega}_{ki}~\forall~kl$\\ \hline\hline
               $\alpha$ &4& (-1,-1) & (-1,-1)  & (-1,-1) & -1  & -1  & 1 \\  
               $\beta$ &2& (-1,-1) & (-1,-1)  & (-1,-1) & \ 1 & \ 1 & 1\\ 
               $\gamma$ &2& ( 1, 1) & ( 1, 1)  & (-1,-1) & \ 1 & \ 1 & 1\\  \hline    
               $1$ &2& ( 1, 1) & ( 1, 1)  & (-1,-1) & -1 & -1 & 1\\        
               $2$ &2& (-1,-1) & ( 1, 1)  & ( 1, 1) & -1 & -1 & 1\\ 
               $3$ &2& ( 1,-1) & ( 1,-1)  & (-1, 1) & -1 & -1 & 1\\ 
               $4$ &2& ( 1,-1) & (-1, 1)  & ( 1,-1) & -1 & -1 & 1\\ 
               \hline\hline
\end{tabular}
\end{center}
\caption{Stacks, wrapping numbers, and torsion charges for a Pati-Salam model. With the choice of structure parameters 
$x_a = \sqrt{3}, x_b =  x_c = x_d = \sqrt{3}/3$, $N=1$ SUSY will be preserved.  The cycles wrapped by each of the stacks pass through the same set of fixed points.}
\label{stacksPS}
\end{table}

\begin{table}[f]
\begin{center}
\begin{tabular}{|c|c|c|c|c|c|c|c|c|c|c|c|c|c|c|c|c|c|c|c|c|} \hline
                 &$\alpha$&$\beta$&$\gamma$&$1$&$2$&$3$&$4$&$\alpha'$&$\beta'$&$\gamma'$&$1'$&$2'$&$3'$&$4'$ \\ \hline\hline
               
$\alpha$   &0& 0 & 0 & 0 & 0 &  0 & 0 & -8  & 4  &-4  & 0  & 0  & 0  & 0 \\ 
$\beta$ &-& 0 & 0 & 0 & 0 & -4 &-4 &  0  &-8  & 0  & 0  & 0  & 0  & 0 \\   
$\gamma$     &-& - & 0 & 0 & 0 &  4 &-4 &  0  & 0  &-8  & 0  & 0  & 0  & 0 \\     
$1$          &-& - & - & 0 & 0 & -8 & 0 &  0  &-4  & 4  &-8  & 0  & 0  & 0 \\         
$2$          &-& - & - & - & 0 &  0 & 0 &  0  &-4  &-4  & 0  &-8  & 0  & 0 \\ 
$3$          &-& - & - & - & - &  0 & 0 &  0  & 0  & 0  & 0  & 0  & 8  & 0 \\  
$4$          &-& - & - & - & - &  - & 0 &  0  & 0  & 0  & 0  & 0  & 0  & 8 \\
\hline\hline

\end{tabular}
\end{center}
\caption{Intersection numbers between different stacks giving rise to fermions in the bifundamental representation. The resulting gauge group and chiral matter content is that of a four-generation Pati-Salam model.}
\label{intnumPS}
\end{table}

\begin{table}[f]
\begin{center}
\begin{tabular}{|c|c|c|} \hline
Stack & Antisymmetric & Symmetric \\ \hline \hline
$\alpha$       & 8             & 0         \\ 
$\beta$        & 8             & 0         \\ 
$\gamma$       & 8             & 0         \\ 
$1$            & 8             & 0         \\ 
$2$            & 8             & 0         \\ 
$3$            &-8             & 0         \\ 
$4$            &-8             & 0         \\ 
\hline\hline
\end{tabular}
\end{center}
\caption{Intersection numbers between different stacks and their images giving
rise to antisymmetric and symmetric representations for a Pati-Salam model.}
\label{ASchiralmatterPS}
\end{table}   

\begin{table}[f]
\begin{center}
\normalsize
\begin{tabular}{|@{}c@{}|@{}c@{}|@{}c@{}|@{}c@{}|@{}c@{}|@{}c@{}|@{}c@{}|@{}c@{}|@{}c@{}|
@{}c@{}|@{}c@{}|}\hline

  Rep. & Multi. &$U(1)_{\alpha}$&$U(1)_{\beta}$& $U(1)_{\gamma}$&$U(1)_1$& $U(1)_2$& $U(1)_3$&$U(1)_4$ & Field \\
\hline \hline
$(\mathbf{4}_{\alpha'} ,\mathbf{2}_{\gamma})$                   & 4 & 1 & 0 & 1 & 0 & 0 & 0 & 0 & Matter\\
$(\mathbf{\bar{4}}_{\alpha'} ,\mathbf{\bar{2}}_{\beta})$        & 4 &-1 &-1 & 0 & 0 & 0 & 0 & 0 & Matter\\
$(\mathbf{2}_{\beta} ,\mathbf{\bar{2}}_{\gamma})^{\star}$       & - & 0 & 1 &-1 & 0 & 0 & 0 & 0 & EW Higgs\\
$(\mathbf{\bar{4}}_{\alpha} ,\mathbf{2}_{\gamma})^{\star}$      & - &-1 & 0 & 1 & 0 & 0 & 0 & 0 & GUT Higgs\\
$(\mathbf{4}_{\alpha} , \mathbf{\bar{2}}_{\beta})^{\star}$      & - & 1 &-1 & 0 & 0 & 0 & 0 & 0 & GUT Higgs\\ \hline
$(\mathbf{6}_{\alpha'\alpha})$                                  & 8 & 2 & 0 & 0 & 0 & 0 & 0 & 0 & -\\
$(\mathbf{1}_{\beta'\beta})$                                    & 8 & 0 & 2 & 0 & 0 & 0 & 0 & 0 & $\phi_{\beta\beta}$\\
$(\mathbf{1}_{\gamma'\gamma})$                                  & 8 & 0 & 0 & 2 & 0 & 0 & 0 & 0 & $\phi_{\gamma\gamma}$\\
\hline\hline
\end{tabular}
\label{spectrumPS}
\caption{The \lq observable\rq \ spectrum of $SU(4)\times SU(2)_L \times SU(2)_R \times [U(2)^4\times U(1)^3]$.  The
$\star'd$ representations indicate light, non-chiral matter which is present between pairs of fractional branes which wrap homologically identical bulk cycles, but differ in their twisted cycles.}
\end{center}
\end{table} 
For Pati-Salam models constructed from bulk D-branes wrapping non-rigid cycles, the gauge symmetry may be broken
to the MSSM by the process of brane splitting, which corresponds to assigning a VEV to an adjoint scalar in the field
theoretic description.  However, this option is not available in the present construction since the adjoint
fields have been eliminated due to the rigidization of the cycles.   

Although the adjoint fields have been eliminated by splitting the bulk D-branes into their fractional consituents, light non-chiral matter in the
bifundamental representation may still appear between pairs of fractional branes~\cite{Blumenhagen:2005tn}. 
These non-chiral states smoothly connect the configuration of fractional D-branes to one consisting of non-rigid D-branes. 
In the present case, all of the fractional D-branes are wrapping bulk cycles which are homologically identical, but differ in their twisted cycles. As discussed in~\cite{Blumenhagen:2005tn}, one may compute the overlap between two such boundary states:
\begin{equation}
\tilde{A}_{a_ia_j} = \int_0^\infty dl\left\langle a_i\right|e^{-2\pi l H_{cl}}\left|a_j\right\rangle +
\int_0^\infty dl\left\langle a_j\right|e^{-2\pi l H_{cl}}\left|a_i\right\rangle.
\end{equation}
Due to the different signs for the twisted sector, it is found
that in the loop channel amplitude
\begin{equation}
A_{a_ia_j} = \int_0^\infty \frac{dl}{l}Tr_{ij+ji} \left(\frac{1+\theta+\omega+\theta\omega}{4}e^{-2\pi l H_{cl}} \right)
\end{equation}
one massless hypermultiplet appears. Thus, the required states to play the role of the Higgs fields are present in this non-chiral sector.  

In principle, one should determine that there are flat 
directions that can give the necessary VEV's to these states. This process would correspond geometrically to a particular brane 
recombination, where the CFT techniques fail and only a field theory 
analysis of D- and F-flat directions is applicable. For instance, a configuration of fractional branes in which one of these states receives a VEV should smoothly connect this configuration to one in which there is a stack of bulk D-branes wrapping a non-rigid cycle that has been split by assigning a VEV to an ajoint scalar.  Such computations are technically very involved and beyond the scope of the present work, and we defer this for later work. 

In Tables 6-9, we present an MSSM model which is obtained from the above Pati-Salam model by separating the stacks as
\begin{eqnarray}
\alpha \rightarrow \alpha_B + \alpha_L, \ \ \ \ \ \beta \rightarrow \beta_{r1} + \beta_{r2}.
\end{eqnarray}     
This does not mean that the stacks are located at different points in the internal space.  After all, there are no adjoint scalars which may receive a VEV.  Rather, this separation reflects that there has been a spontaneous breaking of the Pati-Salam gauge symmetry down to the MSSM by the Higgs mechanism, where we have identified the Higgs states with  $(\mathbf{4},\mathbf{2},1)$ and $(\mathbf{\bar{4}},1, \mathbf{2})$ representations of $SU(4)\times SU(2)_L \times SU(2)_R$ present in the non-chiral sector.    
The resulting gauge group of the model is then given by $SU(3)\times SU(2)_L \times U(1)_{Y}\times SU(2)^4 \times U(1)^8$, and the MSSM hypercharge is found to be
\begin{eqnarray}
Q_Y = \frac{1}{6}\left(U(1)_{\alpha_B} - 3U(1)_{\alpha_L} - 3U(1)_{\beta_{r1}} + 3U(1)_{\beta_{r2}}\right).
\label{hypercharge}
\end{eqnarray}
Of course, this is just
\begin{eqnarray}
Q_Y = \frac{Q_B - Q_L}{2}+ Q_{I_{3R}},
\end{eqnarray} 
where $Q_B$ and $Q_L$ are baryon number and lepton number respectively, while $Q_{I_{3R}}$ is like the third
component of right-handed weak isospin.
\begin{table}[f]
\begin{center}
\begin{tabular}{|c|c|c|c|c|c|c|c|c|} \hline
                 Stack & N & $(n_1,m_1)$&($n_2,m_2)$&$(n_3,m_3)$ & $\epsilon^{\theta}_{ij}~\forall~ij$ &  $\epsilon^{\omega}_{jk}~\forall~jk$ & $\epsilon^{\theta\omega}_{ki}~\forall~kl$\\ \hline\hline
               $\alpha_B$ &3& (-1,-1) & (-1,-1)  & (-1,-1) & -1 & -1 & 1\\ 
               $\alpha_L$ &1& (-1,-1) & (-1,-1)  & (-1,-1) & -1 & -1 & 1\\ 
               $\beta_{r1}$ &1& (-1, -1) & (-1, -1)  & (-1, -1) & \ 1 & \ 1 & 1\\ 
               $\beta_{r2}$ &1& (-1, -1) & (-1, -1)  & (-1, -1) & \ 1 & \ 1 & 1\\
               $\gamma$ &2& ( 1, 1) & ( 1, 1)  & (-1,-1) & \ 1 & \ 1 & 1\\
                \hline      
               $1$ &2& ( 1, 1) & ( 1, 1)  & (-1,-1) & -1 & -1 & 1\\        
               $2$ &2& (-1,-1) & ( 1, 1)  & ( 1, 1) & -1 & -1 & 1\\ 
               $3$ &2& ( 1,-1) & ( 1,-1)  & (-1, 1) & -1 & -1 & 1\\ 
               $4$ &2& ( 1,-1) & (-1, 1)  & ( 1,-1) & -1 & -1 & 1\\ 
               \hline\hline
\end{tabular}
\end{center}
\caption{Stacks, wrapping numbers, and torsion charges for an MSSM-like model. The three-cycles wrapped by each of the stacks pass through the same set of fixed points.}
\label{stacks}
\end{table}

As discussed, up to twelve $U(1)$ factors may obtain a mass \textit{via} the GS mechanism. In order for the hypercharge to remain massless, it must be orthogonal to each of these factors. In this case, there are only four due to the degeneracy of the stacks. These $U(1)$'s remain
to all orders as global symmetries and are given by 
\begin{eqnarray}
U(1)_A  = 3U(1)_{\alpha_B} - U(1)_{\alpha_L}  - U(1)_{\beta_{r1}} - U(1)_{\beta_{r2}} + 2U(1)_{\gamma}-2U(1)_1 \\ \nonumber
+ 2U(1)_2 + 2U(1)_3 + 2U(1)_4, \\ \nonumber \\ \nonumber
U(1)_B  = -3U(1)_{\alpha_B} + U(1)_{\alpha_L} - U(1)_{\beta_{r1}} - U(1)_{\beta_{r2}}+ 2U(1)_{\gamma} + 2U(1)_1            \\ \nonumber + 2U(1)_2  - 2U(1)_3 + 2U(1)_4,   
         \\ \nonumber \\ \nonumber
U(1)_C  = 3U(1)_{\alpha_B} - U(1)_{\alpha_L} - U(1)_{\beta_{r1}} - U(1)_{\beta_{r2}} - 2U(1)_{\gamma} + 2U(1)_1   \\ \nonumber - 2U(1)_2  - 2U(1)_3 + 2U(1)_4,  
         \\ \nonumber \\ \nonumber
U(1)_D  = -3U(1)_{\alpha_B} + U(1)_{\alpha_L} -  U(1)_{\beta_{r1}} - U(1)_{\beta_{r2}}- 2U(1)_{\gamma} - 2U(1)_1 \\ \nonumber  - 2U(1)_2+ 2U(1)_3  + 2U(1)_4.
\label{globalsym}
\end{eqnarray}
Note that the hypercharge orthogonal to each of these $U(1)$ factors and so will remain massless. 
The \lq observable\rq \ sector basically consists of
a four-generation MSSM plus right-handed neutrinos.  The rest of the spectrum primarily consists of vector-like matter, many of which are singlets under
the MSSM gauge group. 
\begin{table}[f]
\begin{center}
\begin{tabular}{|c|c|c|c|c|c|c|c|c|c|c|c|c|c|c|c|c|c|c|c|c|} \hline
                 &$\alpha_B$&$\alpha_L$&$\beta_{r1}$&$\beta_{r2}$&$\gamma$&$1$&$2$&$3$&$4$&$\alpha_B'$&$\alpha_L'$&$\beta_{r1}'$&$\beta_{r2}'$&$\gamma'$&$1'$&$2'$&$3'$&$4'$ \\ \hline\hline
               
$\alpha_B$   &0& 0 & 0 & 0 & 0 & 0 & 0 &  0 & 0 & -8  &-8  & 4  & 4  &-4  & 0  & 0  & 0  & 0 \\ 
$\alpha_L$   &-& 0 & 0 & 0 & 0 & 0 & 0 &  0 & 0 &  0  &-8  & 4  & 4  &-4  & 0  & 0  & 0  & 0 \\ 
$\beta_{r1}$ &-& - & 0 & 0 & 0 & 0 & 0 & -4 &-4 &  0  & 0  &-8  &-8  & 0  & 0  & 0  & 0  & 0 \\  
$\beta_{r2}$ &-& - & - & 0 & 0 & 0 & 0 & -4 &-4 &  0  & 0  & 0  &-8  & 0  & 0  & 0  & 0  & 0 \\ 
$\gamma$     &-& - & - & - & 0 & 0 & 0 &  4 &-4 &  0  & 0  & 0  & 0  &-8  & 0  & 0  & 0  & 0 \\     
$1$          &-& - & - & - & - & 0 & 0 & -8 & 0 &  0  & 0  &-4  &-4  & 4  &-8  & 0  & 0  & 0 \\         
$2$          &-& - & - & - & - & - & 0 &  0 & 0 &  0  & 0  &-4  &-4  &-4  & 0  &-8  & 0  & 0 \\ 
$3$          &-& - & - & - & - & - & - &  0 & 0 &  0  & 0  & 0  & 0  & 0  & 0  & 0  & 8  & 0 \\  
$4$          &-& - & - & - & - & - & - &  - & 0 &  0  & 0  & 0  & 0  & 0  & 0  & 0  & 0  & 8 \\
\hline\hline
                              
\end{tabular}
\end{center}
\caption{Intersection numbers between different stacks giving rise to fermions in the bifundamental representation. The resulting gauge group and chiral matter content is that of a four-generation MSSM-like model.}
\label{intnum}
\end{table}
\begin{table}[f]
\begin{center}
\begin{tabular}{|c|c|c|} \hline
Stack & Antisymmetric & Symmetric \\ \hline \hline
$\alpha_B$     & 8             & 0         \\ 
$\alpha_L$     & 8             & 0         \\ 
$\beta_{r1}$   & 8             & 0         \\ 
$\beta_{r2}$   & 8             & 0         \\ 
$\gamma$       & 8             & 0         \\ 
$1$            & 8             & 0         \\ 
$2$            & 8             & 0         \\ 
$3$            &-8             & 0         \\ 
$4$            &-8             & 0         \\ 
\hline\hline
\end{tabular}
\end{center}
\caption{Intersersection numbers between different stacks and their images giving
rise to antisymmetric and symmetric representations for an MSSM-like model.}
\label{ASchiralmatter}
\end{table}  
\begin{table}[ht]
\begin{center}
\begin{tabular}{|@{}c@{}|@{}c@{}|@{}c@{}|@{}c@{}|@{}c@{}|@{}c@{}|@{}c@{}||@{}c@{}||@{}c@{}|
@{}c@{}|@{}c@{}||@{}c@{}||@{}c@{}|@{}c@{}|@{}c@{}|@{}c@{}||@{}c@{}||@{}c@{}|}\hline

  Rep. & Multi. &$U(1)_{\alpha_B}$&$U(1)_{\alpha_L}$& $U(1)_{\beta_{r1}}$ & $U(1)_{\beta_{r2}}$ & $U(1)_{\gamma}$ & $Q_Y$ & $U(1)_A$ & $U(1)_B$ & $U(1)_C$ & $U(1)_D$ & Field  \\
\hline \hline

$(\mathbf{3}_{\alpha_B'} ,\mathbf{2}_{\gamma})$ & 4 & 1 & 0 & 0 & 0 & 1 & 1/6  &  5 &  -1& 1& -5 & $Q$\\

$(\mathbf{\bar{3}}_{\alpha_B'} ,\mathbf{1}_{\beta_{r2}})$ & 4 & -1 &  0 & 0 & -1 & 0 & -2/3  & -2 & 4&-2 & 4& $U^c$\\

$(\mathbf{\bar{3}}_{\alpha_B'} ,\mathbf{1}_{\beta_{r1}})$ & 4 & -1 &  0& -1 & 0 & 0 &  1/3 &  -2 & 4& -2& 4& $D^c$\\

$(\mathbf{1}_{\alpha_L'} ,\mathbf{2}_{\gamma})$ & 4 & 0 & 1 & 0 & 0 & 1 &  -1/2 & 3 & 1& -1& -3 & $L$ \\

$(\mathbf{1}_{\alpha_L'},\mathbf{1}_{\beta_{r1}})$ & 4 & 0 & -1 & -1 & 0 & 0 & 1 & 0 & 2& 0& 2 &  $E^c$\\

$(\mathbf{1}_{\alpha_L},\mathbf{1}_{\beta_{r2}})$ & 4 & 0 & -1 & 0 &  -1 & 0 & 0 & 0 & 2& 0& 2&  $N$ \\ \hline

$(\mathbf{1}_{\beta_{r1}} ,\mathbf{\bar{2}}_{\gamma})^{\star}$ & - & 0 & 0 & 1 & 0 & -1 & -1/2 & -3 & -3 & 1 & 1 & $H_d$ \\

$(\mathbf{\bar{2}}_{\gamma} ,\mathbf{1}_{\beta_{r2}})^{\star}$ & -  & 0 & 0 & 0 & 1 & -1 & 1/2 & -3 & -3 & 1 & 1 & $H_u$ \\ \hline \hline

$(\mathbf{1}_{\gamma'\gamma})$ & 8 &  0 & 0 & 0 & 0 & -2 & 0 & -4 & -4 & 4& 2& $\phi_{\gamma\gamma}$\\ 

$(\mathbf{1}_{\beta_{r1}'},\mathbf{1}_{\beta_{r2}})$ & 8 &  0 & 0 & 1 & 1 & 0 & 0 & -2 & -2& -2&-2 &  $\phi_{\beta_{r1r2}}$\\

$(\mathbf{3}_{\alpha_B'} ,\mathbf{1}_{\alpha_L})$ & 8 & 1 & 1 & 0 & 0 & 0 & -1/3 & 4 & -4& 4&-4 & $D_1$ \\

$(\mathbf{\bar{3}}_{\alpha_B'\alpha_B})$ & 8 & 2 & 0 & 0 & 0 & 0 & 1/3 & 6 & -6& 6 & -6& $D^c_2$\\
\hline\hline
\end{tabular}
\label{MSSMspectrum}
\caption{The \lq observable\rq \ spectrum of $\left[SU(3)\times SU(2)_L \times U(1)_Y\right]\times U(2)^4 \times U(1)^4$.  The
$\star'd$ representations indicate light, non-chiral matter which exist between pairs of fractional branes which wrap identical bulk cycles, but differ in their twisted cycles.}
\end{center}
\end{table} 
Using the states listed in Table 9, we may construct all of the required MSSM Yukawa couplings,    
\begin{equation}
W_Y = y_u H_u Q U^c + y_d H_d Q D^c + y_l H_d L E^c
\end{equation}
keeping in mind that all of the MSSM fields are charged under the global symmetries defined in eqns. 45. Typically, this results in the
forbidding of some if not all of the desired Yukawa couplings.  In this case, all of the Yukawa couplings are allowed  by the global symmetries including a trilinear Dirac mass term for neutrinos,
\begin{equation}
W_D = \lambda_{\nu}L N H_u.
\end{equation}  
By itself, such a term would imply neutrino masses of the order of the quarks and charged leptons.  However, if in addition there exist 
a Majorana mass term for the right-handed neutrinos,
\begin{equation}
W_m = M_m N N,
\end{equation}
a see-saw mechanism may be employed. Such a mass term
may in principle be generated by $E2$ instanton effects~\cite{Ibanez:2006da, Blumenhagen:2006xt}. This mechanism may also be employed to generate a $\mu$-term of the order of the EW scale.

In addition to the matter spectrum charged under the MSSM gauge groups and total gauge singlets, there is additional vector-like matter transforming under the \lq hidden\rq \ gauge group  $U(2)_1 \otimes U(2)_2 \otimes U(2)_3 \otimes U(2)_4$.  By choosing appropriate flat directions, we may deform the fractional cycles wrapped by these stacks into bulk cycles such that
\begin{equation}
U(2)_1 \otimes U(2)_2 \rightarrow U(1); \ \ \ \ \ U(2)_3 \otimes U(2)_4 \rightarrow U(1).
\end{equation}
Thus, matter transforming under these gauge groups becomes a total gauge singlet or becomes massive and disappears from the spectrum altogether. The remaining eight pairs of exotic color triplets present in the model resulting from the breaking $\mathbf{6}\rightarrow \mathbf{3}\oplus \mathbf{\bar{3}}$, while not truly vector-like due to their different charges under the global symmetries, may in principle become massive via instanton effects in much the same way a 
$\mu$-term may be generated.  

\section{Gauge Coupling Unification}
The MSSM predicts the unification of the three gauge couplings at an energy $\sim2\times10^{16}$~GeV.
In intersecting D-brane models, the gauge groups arise from different stacks of branes, and so
they will not generally have the same volume in the compactified space.  Thus, the gauge couplings
are not automatically unified.  

The low-energy $N=1$ supergravity action is basically determined by the K\"ahler potential $K$, the superpotential
$W$ and the gauge kinetic function $f$.  All of these functions depend on the background space moduli fields.   
\noindent
For branes wrapping cycles not invariant under $\Omega R$, the holomorphic gauge kinetic function for a D6 brane
stack $P$ is given by~\cite{Blumenhagen:2006ci}
\begin{eqnarray}
f_P = \frac{1}{2\pi l_s}\left[e^{\phi}\int_{\Pi_P}\mbox{Re}(e^{-i\theta_P}\Omega_3)-i\int_{\Pi_P}C_3\right]
\end{eqnarray}
from which it follows\footnote{This is closely related to the SUSY conditions.} (with $\theta_P = 0$ for $\Z_2\times\Z_2$)
\begin{eqnarray}
f_P &=& 
(n_P^1\,n_P^2\,n_P^3\,s-n_P^1\,m_P^2\,m_P^3\,u^1-n_P^2\,m_P^1\,m_P^3\,u^2-
n_P^3\,m_P^1\,m_P^2\,u^3)
\label{gaugefunction}
\end{eqnarray}
where $u^i$ and $s$ are the complex structure moduli and dilaton in the field theory basis.
The gauge coupling constant associated with a stack P is given by
\begin{eqnarray}
g_{D6_P}^{-2} &=& |\mathrm{Re}\,(f_P)|.\label{idb:eq:gkf}
\end{eqnarray}
\noindent 
Thus, we identify the $SU(3)$ holomorphic gauge function with stack $\alpha_{B}$, and the $SU(2)$ holomorphic gauge function with stack $\gamma$. The $U(1)_Y$ holomorphic gauge function is then given by taking a linear combination of the holomorphic gauge functions from all the stacks.  
In this way, it is found~\cite{Blumenhagen:2003jy} that
\begin{equation}
f_Y = \frac{1}{6}f_{\alpha_B} + \frac{1}{2}f_{\alpha_L} + \frac{1}{2}f_{\beta_{r1}} + \frac{1}{2}f_{\beta_{r2}}.
\end{equation}
Thus, it follows that the tree-level MSSM gauge couplings will be unified at the string scale
\begin{equation}
g^2_{s} = g^2_{w} = \frac{5}{3}g^2_Y
\end{equation}
since each stack will have the same gauge kinetic function.  

\section{Conclusion}
In this letter, we have constructed an intersecting D-brane model on the $\Z_2 \times \Z_2'$ orientifold background, also known as the $\Z_2 \times \Z_2$ orientifold with discrete torsion, where the D-branes wrap rigid cycles, thus eliminating the extra adjoint matter.  The model constructed is a supersymmetric four generation MSSM-like model obtained from a spontaneously broken Pati-Salam, with a minimum of extra matter. All of the required Yukawa couplings are allowed by global symmetries which arise via a generalized Green-Schwarz mechanism.  In addition, we find that the tree-level gauge couplings are unified at the string scale with a canonical normalization.

The main drawback of this model is that there are four generations of MSSM matter.  However, the existence of a possible fourth generation is rather tightly constrained, although it is not completely ruled out. Of course, the actual fermion masses await a detailed analysis of the Yukawa couplings.  The emergence of three light generations may in fact be correlated with the existence of three twisted sectors. If there turns out to be a fourth generation, then
it would almost certainly be discovered at LHC within the next few years.      
Another interesting possibility is that the presence of discrete torsion will complexify the Yukawa couplings and thereby introduce $CP$ violation into the CKM matrix~\cite{Abel:2002az}. Clearly, there is much work to be done to work out the detailed phenomenology of this model and we plan to return to this topic in the near future.  With the LHC era just around the corner, it would be nice to have testable string models in hand.

\section{Acknowledgements}
The work of C-M Chen is supported by the Mitchell-Heep Chair in High Energy Physics.  The work of D.V. Nanopoulos is supported by DOE grant DE-FG03-95-Er-40917.  We thank
Tianjun Li for a critical reading of the manuscript and for helpful suggestions.  We would also like to thank Mirjam Cvetic for useful discussions and helpful advice.


\newpage

\begin{thebibliography}{99}

\bibitem{Blumenhagen:2004xx}
  R.~Blumenhagen, F.~Gmeiner, G.~Honecker, D.~Lust and T.~Weigand,
  Nucl.\ Phys.\  B {\bf 713}, 83 (2005)
  [arXiv:hep-th/0411173].

\bibitem{Gmeiner:2005vz}
  F.~Gmeiner, R.~Blumenhagen, G.~Honecker, D.~Lust and T.~Weigand,
  JHEP {\bf 0601}, 004 (2006)
  [arXiv:hep-th/0510170].
  
\bibitem{Douglas:2006xy}
  M.~R.~Douglas and W.~Taylor,
  arXiv:hep-th/0606109.

\bibitem{Gmeiner:2006qw}
  F.~Gmeiner,
  arXiv:hep-th/0608227.
  
\bibitem{Gmeiner:2006vb}
  F.~Gmeiner and M.~Stein,
  Phys.\ Rev.\ D {\bf 73}, 126008 (2006)
  [arXiv:hep-th/0603019].
  
\bibitem{Dienes:2006ut}
  K.~R.~Dienes,
  Phys.\ Rev.\ D {\bf 73}, 106010 (2006)
  [arXiv:hep-th/0602286].

\bibitem{Berkooz:1996km}
  M.~Berkooz, M.~R.~Douglas and R.~G.~Leigh,
  Nucl.\ Phys.\ B {\bf 480}, 265 (1996)
  [arXiv:hep-th/9606139].

\bibitem{Bachas:1995ik}
  C.~Bachas,
  [arXiv:hep-th/9503030].

\bibitem{Blumenhagen:2000wh}
  R.~Blumenhagen, L.~G\"orlich, B.~K\"ors and D.~L\"ust,
  JHEP {\bf 0010}, 006 (2000)
  [arXiv:hep-th/0007024];
  R.~Blumenhagen, B.~K\"ors and D.~L\"ust,
  JHEP {\bf 0102}, 030 (2001)
  [arXiv:hep-th/0012156].

\bibitem{Aldazabal:2000dg}
  G.~Aldazabal, S.~Franco, L.~E.~Ib\'a\~nez, R.~Rabad\'an and A.~M.~Uranga,
  J.\ Math.\ Phys.\  {\bf 42}, 3103 (2001)
  [arXiv:hep-th/0011073];
  JHEP {\bf 0102}, 047 (2001)
  [arXiv:hep-ph/0011132].

\bibitem{Angelantonj:2000hi}
  C.~Angelantonj, I.~Antoniadis, E.~Dudas and A.~Sagnotti,
  Phys.\ Lett.\ B {\bf 489}, 223 (2000)
  [arXiv:hep-th/0007090].

\bibitem{Ellis:2002ci}
  J.~R.~Ellis, P.~Kanti and D.~V.~Nanopoulos,
  Nucl.\ Phys.\ B {\bf 647}, 235 (2002)
  [arXiv:hep-th/0206087].
  
\bibitem{CveticShiuUranga}
  M.~Cveti\v c, G.~Shiu and A.~M.~Uranga,
  Nucl.\ Phys.\ B {\bf 615}, 3 (2001)
  [arXiv:hep-th/0107166]. 
    
\bibitem{Cvetic:2001tj}
  M.~Cveti\v c, G.~Shiu and A.~M.~Uranga,
  Phys.\ Rev.\ Lett.\  {\bf 87}, 201801 (2001)
  [arXiv:hep-th/0107143].

\bibitem{Cvetic:P:S:L}
  M.~Cveti\v c and I.~Papadimitriou,
  Phys.\ Rev.\ D {\bf 67}, 126006 (2003)
  [arXiv:hep-th/0303197];
  M.~Cveti\v c, T.~Li and T.~Liu,
  Nucl.\ Phys.\ B {\bf 698}, 163 (2004)
  [arXiv:hep-th/0403061].

\bibitem{Cvetic:2002pj}
  M.~Cveti\v c, I.~Papadimitriou and G.~Shiu,
  Nucl.\ Phys.\ B {\bf 659}, 193 (2003)
  [Erratum-ibid.\ B {\bf 696}, 298 (2004)]
  [arXiv:hep-th/0212177].
  
\bibitem{Cvetic:2004nk}
  M.~Cveti\v c, P.~Langacker, T.~Li and T.~Liu,
  Nucl.\ Phys.\ B {\bf 709}, 241 (2005)
  [arXiv:hep-th/0407178].
 
\bibitem{Chen:2006gd}
  C.~M.~Chen, T.~Li and D.~V.~Nanopoulos,
  Nucl.\ Phys.\ B {\bf 740}, 79 (2006)
  [arXiv:hep-th/0601064].
  
\bibitem{Barr:1981qv}
  S.~M.~Barr,
  Phys.\ Lett.\ B {\bf 112}, 219 (1982);
  Phys.\ Rev.\ D {\bf 40}, 2457 (1989).

\bibitem{FSU(5)N}
  J.~P.~Derendinger, J.~E.~Kim and D.~V.~Nanopoulos,
  Phys.\ Lett.\ B {\bf 139}, 170 (1984);

\bibitem{AEHN}
I.~Antoniadis, J.~R.~Ellis, J.~S.~Hagelin and D.~V.~Nanopoulos,
Phys.\ Lett.\ B {\bf 194} (1987) 231;
Phys.\ Lett.\ B {\bf 205} (1988) 459;
Phys.\ Lett.\ B {\bf 208} (1988) 209
[Addendum-ibid.\ B {\bf 213} (1988) 562];
Phys.\ Lett.\ B {\bf 231} (1989) 65.
   
\bibitem{Chen:2005ab}
  C.~M.~Chen, G.~V.~Kraniotis, V.~E.~Mayes, D.~V.~Nanopoulos and J.~W.~Walker,
  Phys.\ Lett.\ B {\bf 611}, 156 (2005)
  [arXiv:hep-th/0501182].

\bibitem{Axenides:2003hs}
  M.~Axenides, E.~Floratos and C.~Kokorelis,
  JHEP {\bf 0310}, 006 (2003)
  [arXiv:hep-th/0307255].

\bibitem{Chen:2005cf}
  C.~M.~Chen, V.~E.~Mayes and D.~V.~Nanopoulos,
  Phys.\ Lett.\ B {\bf 633}, 618 (2006)
  [arXiv:hep-th/0511135].

\bibitem{Chen:2005mm}
  C.~M.~Chen, G.~V.~Kraniotis, V.~E.~Mayes, D.~V.~Nanopoulos and J.~W.~Walker,
  Phys.\ Lett.\ B {\bf 625}, 96 (2005)
  [arXiv:hep-th/0507232].
  
\bibitem{Chen:2006ip}
  C.~M.~Chen, T.~Li and D.~V.~Nanopoulos,
  [arXiv:hep-th/0604107].

\bibitem{Blumenhagen:2005mu}
  R.~Blumenhagen, M.~Cvetic, P.~Langacker and G.~Shiu,
  Ann.\ Rev.\ Nucl.\ Part.\ Sci.\  {\bf 55}, 71 (2005)
  [arXiv:hep-th/0502005].
  
\bibitem{Blumenhagen:2006ci}
  R.~Blumenhagen, B.~Kors, D.~Lust and S.~Stieberger,
  [arXiv:hep-th/0610327].
  
\bibitem{Dudas:2005jx}
  E.~Dudas and C.~Timirgaziu,
  Nucl.\ Phys.\ B {\bf 716}, 65 (2005)
  [arXiv:hep-th/0502085].
  
\bibitem{Blumenhagen:2005tn}
  R.~Blumenhagen, M.~Cvetic, F.~Marchesano and G.~Shiu,
  JHEP {\bf 0503}, 050 (2005)
  [arXiv:hep-th/0502095].
  
\bibitem{Blumenhagen:2002wn}
  R.~Blumenhagen, V.~Braun, B.~Kors and D.~Lust,
  JHEP {\bf 0207}, 026 (2002)
  [arXiv:hep-th/0206038].
    
\bibitem{Ibanez:2006da}
  L.~E.~Ibanez and A.~M.~Uranga,
  [arXiv:hep-th/0609213].

\bibitem{Blumenhagen:2006xt}
  R.~Blumenhagen, M.~Cvetic and T.~Weigand,
  [arXiv:hep-th/0609191].
 
\bibitem{Cremades:2003qj}
  D.~Cremades, L.~E.~Ibanez and F.~Marchesano,
  JHEP {\bf 0307}, 038 (2003)
  [arXiv:hep-th/0302105].
   
\bibitem{Lust:2004cx}
  D.~Lust, P.~Mayr, R.~Richter and S.~Stieberger,
  Nucl.\ Phys.\ B {\bf 696}, 205 (2004)
  [arXiv:hep-th/0404134].
  
\bibitem{Blumenhagen:2003jy}
  R.~Blumenhagen, D.~Lust and S.~Stieberger,
  JHEP {\bf 0307}, 036 (2003)
  [arXiv:hep-th/0305146].
  
\bibitem{Abel:2002az}
  S.~A.~Abel and A.~W.~Owen,
  Nucl.\ Phys.\ B {\bf 651}, 191 (2003)
  [arXiv:hep-th/0205031].
  
\end{thebibliography}
\end{document}